\documentclass[
aps,%
11pt,%
final,
notitlepage,%
oneside,%
onecolumn,%
nobibnotes,%
nofootinbib,%
preprintnumbers,%
superscriptaddress,%
showpacs,%
centertags,%
showkeys,%
amsmath,%
amssymb]{revtex4}

\usepackage{amsmath,amssymb} 
\usepackage[dvips]{graphicx} 

\newcommand{\m}[1]{\ensuremath{\mathrm {#1}}}
\newcommand*{\dd}{\ensuremath{\mathrm{d}}}

\newcommand*{\Li}{\ensuremath{\mathrm{Li}_2}}

\newcommand\nn{\nonumber}

\begin{document}

\title{Radiative corrections to muon decay in leading and next to
leading approximation for electron spectrum}

\author{E.~Barto\v{s}}
\affiliation{Institute of Physics, Slovak Academy of Sciences, Bratislava}

\author{E.~A.~Kuraev}
\affiliation{Joint Institute for Nuclear Research, Dubna, Russia}

\author{M.~Se{\v c}ansk\'y}
\affiliation{Institute of Physics, Slovak Academy of Sciences, Bratislava}


\begin{abstract}
We have noted that the electron spectrum of muon decay in the leading logarithmic approximation calculated in two lowest orders of the perturbation theory in the paper of Berman (1958), can be reproduced by the parton language. This fact permits one to generalize the result to all orders of the perturbation theory using the structure function method.
\end{abstract}

\pacs{}
\keywords{muon decay}

\maketitle

\section{Introduction}

The lowest order radiative corrections (RC) to the muon weak decay width where calculated about fifty years ago \cite{Berman:1958ti}. The result for the electron spectrum in muon decay including RC was obtained in the form
\begin{equation}
\frac{\dd W^{(1)}(x)}{\dd x} = \frac{\dd W_\m{B}(x)}{\dd x}\Big[1 + \frac{\alpha}{2\pi}h(x)\Big],\quad x=\frac{E_\m{e}}{E_\m{max}},\quad h(x)=A(x)+LB(x),\quad L=\ln\frac{M^2}{m^2}
\end{equation}
with the spectrum in Born approximation
\begin{equation}
\frac{\dd W_\m{B}(x)}{\dd x}=2W_\m{B}, \quad W_\m{B}=\frac{G^2M^5}{192\pi^3}x^2(3-2x),
\end{equation}
here $M$ is the muon mass, $m$ is the electron mass, $L$ is so called ``large logarithm'' ($L\approx 12$). The result of lowest order RC including is presented in the expression $h(x)$, or in the functions $A(x)$ and $B(x)$ respectively \cite{Lifshitz}
\begin{align}
A(x)=&\,4\Li(x)-\frac{2\pi^2}{3}-4+2\Big[3\ln(1-x)-2\ln x+1\Big]\ln x-2\frac{1+x}{x}\ln(1-x)  \nn \\
&+\frac{(1-x)(5+17x-16x^2)}{3x^2(3-2x)}\ln
x+\frac{(1-x)(-22x+34x^2)}{3x^2(3-2x)},\nn \\
B(x)=&3+4\ln\frac{1-x}{x}+\frac{(1-x)(5+17x-34x^2)}{3x^2(3-2x)}.
\end{align}
One must remark that the result of the calculations does not suffer from the ultraviolet and the infrared divergences. Besides it satisfies Kinoshita--Lee--Nauenberg (KLN) theorem  \cite{KLN} about
the cancellation of mass singularities, namely the total width is
finite in the limit of zero electron mass
\begin{equation} \label{eq:intB}
\int\limits_0^1\dd x\, \frac{\dd W_\m{B}(x)}{\dd x}B(x)=0.
\end{equation}

The mechanism of the realization of KLN theorem can be understand from the positions of parton interpretation of Quantum Electrodynamics (QED). Really, one can be convinced in the validity of the relation
\begin{equation}
\frac{1}{2}x^2(3-2x)h(x)= (L-1)\int\limits_x^1 \frac{\dd y}{y}
y^2(3-2y) P(\frac{x}{y})+K(x),\quad K(x)=\frac{1}{2}(A(x)+B(x))x^2(3-2x),
\end{equation}
where
\begin{gather*}
P(z)=\left(\frac{1+z^2}{1-z}\right)_+=\lim_{\Delta \to 0} \left[\frac{1+z^2}{1-z}\theta(1-z-\Delta)+(2\ln\Delta+ \frac{3}{2})\delta(1-z)\right]
\end{gather*}
is the kernel of the evolution equation of twist two operators.
Using the property $\int\limits_0^1 \dd x P(x)=0$, one can validate Eq.~(\ref{eq:intB})
\begin{equation}
W_\m{B}\int\limits_0^1 \dd x\, x^2(3-2x)B(x)=\int\limits_0^1 \dd x\int\limits_x^1 \frac{\dd y}{y} \frac{\dd W_\m{B}(y)}{\dd y}P(\frac{x}{y})= \nn \\
\int\limits_0^1 \dd y \frac{\dd W_\m{B}(y)}{\dd y}\int\limits_0^y
\frac{\dd x}{y}P(\frac{x}{y})=0.
\end{equation}
Considering the process in Born approximation as a ``hard'' process
and applying Collins factorization theorem about the
contributions of the short and long distances one can generalize the lowest order result to include all terms of the sort $(\alpha
L/\pi)^n$ (leading logarithmical approximation (LLA)) as well as
the terms of the sort $\alpha(\alpha L/\pi)^n$ (next to leading
approximation (NLO)) in the form
\begin{equation} \label{eq:W}
\frac{\dd W(x)}{\dd x}= \int\limits_x^1
\frac{\dd y}{y} \frac{\dd W_\m{B}(y)}{dy} D\Big(L,\frac{x}{y}\Big)\Big(1+\frac{\alpha}{\pi}K(y)\Big),
\end{equation}
with
\begin{equation} \label{eq:D}
D(L,z)=\delta(1-z)+\frac{\alpha}{2\pi} (L-1)P(z)+ \frac{1}{2!}\Big(\frac{\alpha}{2\pi}(L-1)\Big)^2P^{(2)}(z)+\dots,
\end{equation}
and
\begin{align}
K(y)=& \,\Big[2\Li(y)-\frac{\pi^2}{3}-\frac{1}{2}+\Big[3\ln(1-y)-2\ln y+1\Big] \ln y -\frac{1+y}{y}\ln(1-y)\nn \\ &+\frac{1-y}{6y^2(3-2y)}\Big[(5+17y-16y^2)\ln y+5(1-y)\Big] + 2\ln\frac{1-y}{y}\Big]y^2(3-2y).
\end{align}

For the numerical calculations one can use for the structure function $D(L,z)$ from Eq.~(\ref{eq:D}) the "smoothed" (but
equivalent) form \cite{Kuraev:1985hb}
\begin{gather}
D(L,z)= \frac{\beta}{2}(1-z)^{\frac{\beta}{2}-1}
\Big(1+ \frac{3}{8}\beta\Big)-\frac{\beta}{4}(1+z)+O(\beta^2),\quad
\beta=\frac{2\alpha}{\pi}(L-1).
\end{gather}
One can find useful relation
\begin{equation} \label{eq:DPhi}
\int\limits_x^1 \frac{\dd y}{y^2}D\Big(L,\frac{x}{y}\Big)\Psi(y) =
\int\limits_x^1 \frac{\dd y}{y^2}D\Big(L,\frac{x}{y}\Big)\Big[\Psi(y)-\Psi(x)\Big] +
\frac{1}{x}\Psi(x) \int\limits_x^1 \dd z D(L,z)
\end{equation}
with
\begin{gather*}
\Psi(y)=y^3(3-2y)\Big(1+\frac{\alpha}{\pi}K(y)\Big),\\
\int\limits_x^1 \dd z D(L,z) = (1-x)^{\frac{\beta}{2}}
\Big(1+ \frac{3}{8}\beta\Big)-\frac{\beta}{8}(1-x)(3+x)+O(\beta^2).
\end{gather*}

\begin{figure} \label{fig:1}
\includegraphics[width=.4\textwidth]{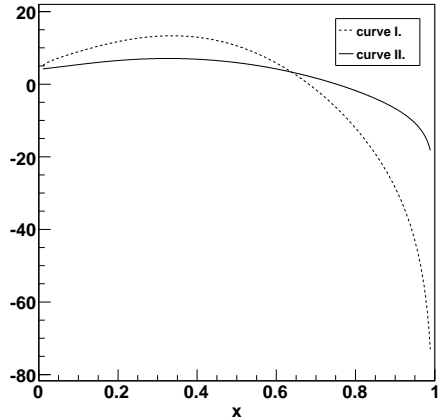}
\caption{The deviation of the electron spectrum in muon decay from the spectrum in Born approximation: for the lowest order (I.), for all orders (II.) of perturbation theory.}
\end{figure}

For the comparison we have given in Fig.~\ref{fig:1} the numerical values of the quantity
$\frac{96\pi^3}{G^2M^5}\Big(\frac{\dd W^{(1)}(x)}{\dd x}-\frac{\dd W_\m{B}(x)}{\dd x}\Big)\frac{\pi}{\alpha}=\frac{1}{2}h(x)x^2(3-2x)$ -- the curve I., and the quantity $\frac{96\pi^3}{G^2M^5}\Big(\frac{\dd W(x)}{\dd x}-\frac{\dd W_\m{B}(x)}{\dd x}\Big)$ -- the curve II., calculated in LLA and NLO approximations (see (\ref{eq:W}),(\ref{eq:DPhi})). One can see that the spectrum contrary to the result of lowest order of perturbation theory is well defined in the whole region of $x$ including $x\to 0$ and $x\to 1$. The total width with RC does not contain ``large logarithm'' due to the property $\int\limits_0^1 \dd z D(L,z)=1$. In our approach we have obtained
\begin{equation}
W=W_\m{B}\Big[1-\frac{\alpha}{2\pi}\Big(\pi^2-\frac{25}{4}\Big)+o(\alpha^2)\Big].
\end{equation}
Let us note, that the terms of order $\alpha^2$ were calculated in \cite{vanRitbergen:1998yd}.

\acknowledgments

One of us (E.A.K.) is grateful to the Institute of Physics, SAS for warm hospitality during the time of his stay. E.A.K. acknowledges the support of INTAS (grant no. 05-1000008-8328). The work was also
supported in part by the Slovak Grant Agency for Sciences
VEGA, Grant No. 2/7116.

\end{document}